\documentclass[aps,prl,twocolumn,showpacs]{revtex4}
\usepackage{amsmath}
\usepackage{graphicx,epsfig,psfrag}
\usepackage{amssymb}

\renewcommand{\vec}[1]{{\bf #1}}
\begin{document}
\title{Metastable superfluidity of repulsive fermionic atoms in optical lattices}

\author{Achim Rosch}
\author{David Rasch}
\author{Benedikt Binz}
\author{Matthias Vojta}
\affiliation{Institute of Theoretical Physics, University of Cologne,
Z\"ulpicher Str. 77, 50937 Cologne, Germany}

\begin{abstract}
  In the fermionic Hubbard model, doubly occupied states have an
  exponentially large lifetime for strong repulsive interactions $U$.
  We show that this property can be used to prepare
  a metastable $s$-wave superfluid state for fermionic atoms in
  optical lattices described by a large-$U$ Hubbard model. When an
  initial band-insulating state is expanded, the doubly occupied sites
  Bose condense. A mapping to the ferromagnetic Heisenberg model in an
  external field allows for a reliable solution of the problem.
  Nearest-neighbor repulsion and pair hopping are important in stabilizing
  superfluidity.
\end{abstract}


\date{Dec 15, 2008}
\maketitle


Trapped cold atoms open the possibility to realize new quantum states of matter and to
control them with an unprecedented precision. An especially exciting perspective is the
possibility to study interacting quantum systems out of equilibrium. The high tunability
in combination with the slow dynamics of cold atoms allows to
investigate time-dependent processes, for example the quench from a superfluid to a Mott
insulating state \cite{greiner}.

Thermal equilibrium is usually dominated by low-energy
states of the system, while out of equilibrium also high-energy states can
become important. In the continuum, high-energy states typically decay
rapidly in the presence of interactions: High energy implies that
the available phase space for inelastic scattering is
large.  In contrast, for lattice systems where the kinetic energy of a
single particle cannot exceed its bandwidth $D$, a state with high
energy, $E \gg D$ (e.g. a doubly occupied site in a strongly repulsive Hubbard
model), cannot easily decay.
This is a consequence of energy conservation:
To dissipate the huge energy $E$, a complex
many-particle scattering process is needed, with at least $n \gtrsim
E/D$ participating particles. For local two-particle interactions such
processes are expected to be exponentially suppressed for large $n$
(see below).
This effect has been directly observed in measurements of the lifetime
of doubly occupied lattices sites for bosonic $^{87}$Rb atoms in an
optical lattice~\cite{Zoller06}:
Starting from a dense cloud of atoms with many doubly occupied sites,
the strength of the trapping potential was reduced in one direction,
allowing the cloud to expand.
Subsequently, many long-lived double occupancies were detected,
with a lifetime exceeding their inverse tunneling
rate by more than two orders of magnitude.

The large lifetime of doubly occupied lattice sites implies that one
can easily create new metastable states of matter. Indeed, numerical
simulations by Kollath {\it et al.} \cite{quench} show that
metastable states form in the
one-dimensional  bosonic Hubbard model for strong repulsion.

An obvious question is whether the doubly occupied sites will Bose condense.
For a bosonic Hubbard model, this question was
investigated  by Petrosyan {\it et al.} \cite{Fleischhauer07}, but the authors
found that instead the system will phase-separate:
Due to nearest-neighbor attractive interactions, doubly occupied sites
will stick together instead of forming a low-density superfluid.
In this paper we will prove that for fermions, in contrast, a
Bose condensate of spin singlets with $s$-wave symmetry will
form.
Interestingly, the many-particle wavefunction of the relevant homogeneous metastable
superfluid state can be constructed in a controlled
way. It has been known for a long time \cite{eta1,eta2}
that a hidden SU(2) symmetry of the charge
sector (called SU$_{\rm C}$(2) in the following) of the Hubbard
model can be employed to build wavefunctions with off-diagonal long-range order
(states with so-called ``$\eta$ pairing''\cite{eta1}).
We shall show that these states can easily be realized just by expanding an
atomic cloud in an optical lattice slowly compared to typical collision times
but rapidly compared to the exponentially large lifetime
of the doubly occupied states.

\begin{figure}
\includegraphics[width= \linewidth]{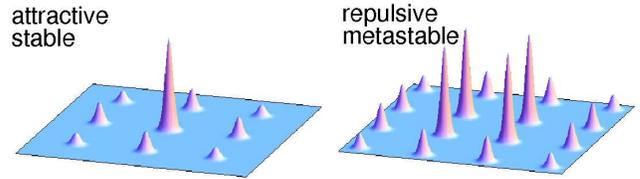}
\caption{Schematic plot of the momentum distribution of fermion
  pairs \cite{regal}. For attractive interactions, the Cooper pairs condense at
  momentum $0$ (and corresponding reciprocal lattice vectors). In contrast, the
  metastable superconductivity of the repulsive Hubbard model arises
  at momentum $(\pm \pi,\pm \pi,\pm \pi)$.
\label{schematic}}
\vspace*{-10pt}
\end{figure}

The condensation of doubly occupied sites can be detected by measuring
the momentum distribution of fermion pairs~\cite{regal}. The
repulsively bound doubly occupied sites of the repulsive Hubbard model
hop from site to site via virtual {\em low-energy} states. Therefore
the sign of their effective hopping amplitude is reversed compared to
bound pairs in the attractive Hubbard model.
This implies that the condensation occurs at momentum
$(\pi,\pi,\pi)$ \cite{eta1,eta2} rather than zero, allowing for an
unambiguous detection of this state, see Fig.~\ref{schematic}.


{\em Setup.}
We consider the fermionic Hubbard model
\begin{eqnarray}\label{H}
\mathcal{H} =
-J \sum_{\langle ij\rangle, \sigma} c^\dagger_{i\sigma}c_{j\sigma} + U \sum_i n_{i \uparrow} n_{i\downarrow}
+V_t \sum_i \vec{r}_i^2 n_i
\end{eqnarray}
on a cubic lattice with a harmonic trapping potential of strength $V_t$.
Here $J$ is the tunneling rate between neighboring sites of the optical lattice,
$U > 12 J=D$ is a strong repulsive interaction, and
$n_{i\sigma}=c^\dagger_{i\sigma}c_{i\sigma}$, $n_i=n_{i \downarrow}+n_{i \uparrow}$.
The lattice distance $a$ is set to unity.

As argued above, the total number of doubly
occupied sites, $N_d=\sum_i n_{i \uparrow} n_{i\downarrow}$, has an extremely long
lifetime, due to the difficulty in loosing the large energy $U$.
This can formally be seen using a well-known unitary transformation \cite{schriefferWolff1,schriefferWolff2},
$\mathcal{H} \to \tilde{\mathcal{H}}=e^{i S} \mathcal{H} e^{-i S}$, $N_d \to \tilde{N}_d=e^{i S} N_d e^{-i S}$,
called Schrieffer-Wolff transformation. For a given arbitrary order $n$, one can explicitly
construct \cite{schriefferWolff2}
a unitary operator $e^{i S_n}$, such that the commutator $[\tilde{N}_d,\tilde{\mathcal{H}}]$
vanishes exactly up to terms of order $1/U^n$.
This proves that,
in the limit of large $U$,
the lifetime $\tau_d$ of doubly occupied sites grows faster than any power of $U$.
The underlying physical reason, the energy bottleneck, has been described
in the introduction. We therefore expect that, for $V_t=0$, $\tau_d$
is exponentially large in $U/D$.


We consider an initial situation where the atoms are densely packed,
with two atoms per site in the center of the trap (i.e. a band insulator state),
and investigate the evolution of the system upon reducing the strength $V_t$
(i.e. curvature) of the trapping potential \cite{alternatively}.
The initial system is in thermal equilibrium, and we assume vanishing entropy for simplicity
(all of the following arguments remain valid as long as the entropy per particle remains small
compared to unity).
The radius of the atomic cloud is $r_d\sim N_d^{1/3}$. To avoid a decay of the doubly occupied
 states by a conversion of interaction energy
  into potential energy, the slope of the trapping
  potential at the edge of the cloud has to be small compared to $U$,
  $2 V_t r_d \ll U$. Taking into account the Mott-insulating shell
  forming around the band-insulating core \cite{Helmes08,Hofstetter08,bloch08}, one obtains
  from this condition the ratio of the numbers of singly and doubly occupied sites,
  $N_1/N_d \gg 1/N_d^{1/3}$.
Nevertheless, the ratio  $N_1/N_d$ can be made sufficiently small, such that singly
occupied states can be neglected.
Note that this is {\it not} required to obtain Bose condensation of double occupancies,
but simplifies the theoretical analysis considerably.

{\em Effective model.}
Neglecting singly occupied sites, the effective Hamiltonian after the Schrieffer-Wolff
transformation \cite{schriefferWolff1,schriefferWolff2} reads \cite{pot}  (up to constant
contributions)
\begin{align}\label{h2}
 \tilde{\mathcal{H}} =
 \frac{J^2}{U} \sum_{\langle ij \rangle}  c^\dagger_{i \uparrow}   c^\dagger_{i \downarrow}
 c_{j \downarrow}   c_{j \uparrow}    + n_{i \uparrow} n_{i \downarrow} (1-n_{j \uparrow})(1- n_{j \downarrow})
 \nonumber \\+
 2 V_t \sum_i \vec{r}_i^2 n_{i\uparrow}n_{i\downarrow}.
 \end{align}
 The first term describes the hopping of doubly occupied sites, the
 second an effective interaction.  In the presence of singly occupied
 sites, the leading correction to (\ref{h2}) arises
 \cite{schriefferWolff2} from
 $J \sum_{\langle ij \rangle \sigma}
 n_{i,-\sigma} c^\dagger_{i \sigma} c_{j \sigma} n_{j,-\sigma}$, which
 describes an exchange of a doubly and a singly occupied site.
This term can be neglected when the local density of single
occupancies is smaller than $J/U$.
While this is not the case at the border of the atomic cloud in its
initial configuration, it turns out to be valid in the scaling limit
discussed below, as single occupancies are efficiently diluted
when the trapping potential gets weaker.


It is useful to rewrite (\ref{h2}) in two different ways. First, one
can identify the doubly occupied states with a boson
$d^\dagger_i=c^\dagger_{i \uparrow} c^\dagger_{i \downarrow}$ such
that (up to a constant)
\begin{align} \label{h3}
 \tilde{\mathcal{H}} =
 \frac{J^2}{U} \sum_{\langle ij \rangle}  d^\dagger_{i} d_{j}    +\sum_{ij} V_{ij} n_{di} n_{d j}+
 2 V_t \sum_i \vec{r}_i^2 n_{di}
 \end{align}
with $n_{di}= d^\dagger_{i}    d_{i}$.
Here $V_{ii}=\infty$ implements a hard-core constraint, and $V_{ij}=-J^2/U$
describes an attraction for nearest neighbors $i$ and $j$.
Second, one can map the hard-core bosons to spins~\cite{Fleischhauer07}.
Starting from (\ref{h2}), this can be
done by performing a particle-hole transformation for the down-spins only, $c^\dagger_{i
\uparrow} \to \tilde{c}^\dagger_{i \uparrow}$, $c^\dagger_{i \downarrow} \to (-1)^i \tilde{c}_{i
\downarrow}$. This maps an empty site to a spin down and a doubly occupied site to a spin
up.
A finite magnetization in the $xy$ plane describes Bose condensation of pairs
of fermions, see below. Identifying
$\vec{S}_i=\frac 1 2 \sum_{\alpha \beta} \tilde{c}^\dagger_{i\alpha} \vec{\sigma}_{\alpha \beta} \tilde{c}_{i \beta}$
after this transformation, one obtains a {\em ferromagnetic} Heisenberg model
in a magnetic field:
\begin{align}\label{h4}
 \tilde{\mathcal{H}} =
 -\frac{J^2}{U} \sum_{\langle ij \rangle} \vec{S}_i \cdot \vec{S}_j  +
 2 V_t \sum_i \vec{r}_i^2 S^z_i.
 \end{align}
The SU(2) symmetry of the first term in (\ref{h4}) is a direct consequence of the
SU$_{\rm C}$(2) symmetry of the charge sector of the underlying Hubbard model
\cite{eta1,eta2}:
For $V_t=0$ and a chemical potential $\mu=U/2$, $\mathcal{H}$ (\ref{H}) commutes
with all three components of the particle-hole transformed operators
$\sum_i \vec{S}_i$ defined above -- in the original variables,
these are $(\eta+\eta^\dagger)/2$, $(\eta-\eta^\dagger)/(2i)$, and $\sum_i(n_i\!-\!1)/2$
with $\eta^\dagger=\sum_i (-1)^i c^\dagger_{i\uparrow} c^\dagger_{i\downarrow}$.

\begin{figure}
\includegraphics[width=0.95 \linewidth,clip]{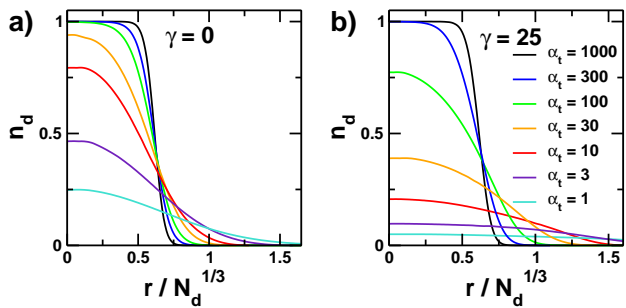}
\caption{
Density profiles of double occupancies
in the trap for different values of the trapping potential $V_t$,
expressed as $\alpha_t=V_t N_d^{4/3} U/J^2$,
a) for vanishing and
b) for finite next-neighbor repulsion and pair hopping,
$\gamma=(6 U'+3 J_p/2) U N_d^{2/3}/J^2$, see text.
Upon reducing $V_t$, the cloud expands and,
simultaneously, the double occupancies condense,
$|\langle c^\dagger_{\uparrow} c^\dagger_\downarrow\rangle|^2/n_d=1-n_d$.
In b), the atomic cloud  spreads out further and superfluidity
is enhanced.
}
\label{densityprofile}
\end{figure}

For $V_t=0$, the exact ground state of (\ref{h4}) for fixed particle density $n_d$ is a ferromagnetic
state,
 \begin{eqnarray}
 | \Psi \rangle = e^{- i \theta \sum_i S^x_i}
 | \uparrow \uparrow \uparrow... \rangle = e^{- i \frac{\theta}{2} \sum_i (-1)^i (c^\dagger_{i \uparrow}
 c^\dagger_{i \downarrow}+ h.c.) } | 0 \rangle
 \end{eqnarray}
 where $\cos \theta=1-2 n_d$ fixes the magnetization
 $S_z=n_d-\frac{1}{2}$ in $z$ direction. The rotational symmetry
 around the $z$ axis is spontaneously broken, which implies
 off-diagonal long-range order, $\langle c^\dagger_{i \uparrow}
 c^\dagger_{i \downarrow} \rangle=\langle d^\dagger_i \rangle=
 \frac{(-1)^i}{2} \sin \theta$, with momentum $(\pi,\pi,\pi)$.  The
 superfluid fraction, defined as $|\langle d^\dagger_i \rangle
 |^2/n_d$, is given exactly by $(1-n_d)$ \cite{gutzwiller}.

The initial thermally equilibrated state, described above, is not
superfluid. A state with a finite expectation value
of $\langle c^\dagger_{i \uparrow} c^\dagger_{i \downarrow} \rangle$
can, however, easily be generated by reducing the trapping potential
$V_t$ adiabatically, i.e., much slower than the typical time scales of
order $U/J^2$ arising from the dynamics of the effective Hamiltonian
(\ref{h2}). If the whole experiment is furthermore performed on a
time scale short compared to the (exponentially large) life time
$\tau_d$ of doubly occupied sites, the system remains in the ground
state of the effective Hamiltonian (\ref{h2}) [or, equivalently,
(\ref{h4})] with the number of doubly occupied sites (the total
magnetization) kept fixed.
To calculate how a finite superfluid fraction arises when $V_t$ is lowered for fixed $N_d$,
we employ a mean-field approximation. We approximate the ground-state wavefunction of
(\ref{h4}) by $|\Psi \rangle = \Pi_i |\hat{n}_i \rangle$, where $|\hat{n}_i \rangle$
describes a spin $i$ polarized in the $+\hat{n}_i$ direction. Here $\hat{n}_i=\hat
n(\vec{r}_i)$ is (in the limit of large $N$ and small $V_t$) a unit vector smoothly
varying as a function of $\vec{r}$, which minimizes
\begin{eqnarray}\label{energy}
E= \int d^3 r
\left[ \frac{J^2}{4 U} (\nabla \hat{n}(\vec{r}))^2+ V_t \vec{r}^2 (\hat{n}^z(\vec{r})+1)\right]
\end{eqnarray}
with the constraint $\int d^3 r [\hat{n}^z(\vec{r})+1]=2 N_d$.
The employed continuum limit becomes formally exact for a large number of atoms
(typical optical-lattice experiments use $10^4$ to $10^5$ atoms).
In this ``scaling'' limit it is convenient to measure distances in units of $N_d^{1/3}$,
and all physical properties only depend on the dimensionless quantity
$\alpha_t=V_t N_d^{4/3} U/J^2$.
The initial state then corresponds to $\alpha_t=\infty$.

\begin{figure}
\begin{center}
\ \ \ \ \
\includegraphics[width=0.7 \linewidth,clip]{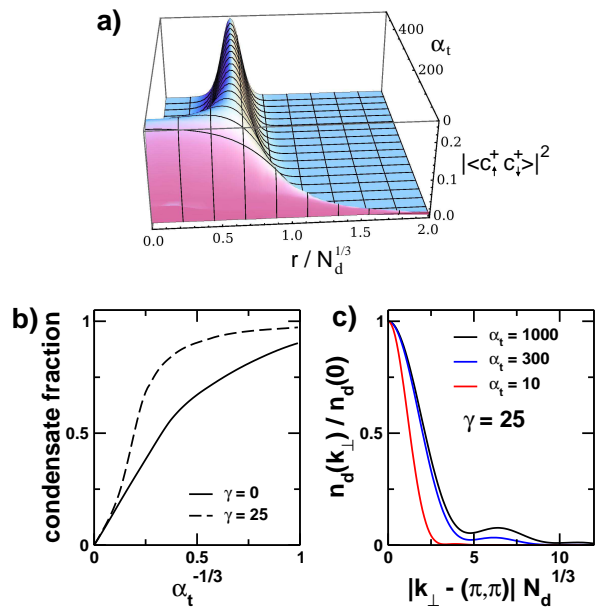}
\includegraphics[width=0.9 \linewidth,clip]{impulsverteilungetc.eps}
\end{center}
\vspace*{-8pt}
\caption{
a)
Square of condensate amplitude,
$|\langle c^\dagger_\uparrow c^\dagger_\downarrow\rangle |^2$,
as function of the radius and the strength of the trapping
potential $\alpha_t$, for next-neighbor repulsion and pair hopping
$\gamma=25$.
b)
Superfluid fraction as function of the trapping potential,
plotted as $\alpha_t^{-1/3}$.
c)
Projected momentum distribution
$n_d(k_\perp) = \int d k_z \langle d^\dagger_k d_k \rangle$,
for different $\alpha_t$ as function of the distance to $(\pi,\pi)$.
The condensate leads to a characteristic peak,
while the non-condensed part gives rise to a uniform background
(not shown).
The oscillations for large $\alpha$ are signatures of the
superfluid shell in the early stage of the expansion.
}
\label{mix}
\end{figure}

{\em Results.}
When the confining potential $V_t$ gets weaker, the cloud of atom pairs
expands (Fig.~\ref{densityprofile}a), and simultaneously a condensate of fermionic pairs first
emerges at the boundary of the band-insulating state, i.e. at the domain wall separating
the spin-up and spin-down phases (Fig.~\ref{mix}a).
When $\alpha_t$ becomes of order $1$ and smaller, the maximum of $|\langle c^\dagger_\uparrow
c^\dagger_\downarrow\rangle|$ moves to the center of the trap, and for $\alpha_t \to 0$
the condensate fraction shown in Fig.~\ref{mix}b approaches unity.
The minimization of (\ref{energy}) then becomes equivalent to solving the Schr\"odinger equation
of non-interacting bosons in a trap using  $\hat{n}\approx
(\Psi(\vec{r}),0,-1+\Psi(\vec{r})^2/2)$.


Is the system a true superfluid?
A superfluid is very different from a ferromagnet,
as the excitation spectrum of the former is linear in momentum while it is quadratic
in the latter.
As the bosonic Hamiltonian (\ref{h2}) is equivalent to a ferromagnetic model,
we conclude that, for $V_t=0$,
the doubly occupied states do {\em not} form a superfluid with a finite phase
stiffness, but only a Bose-Einstein condensate as for non-interacting bosons:
At low energy and density, the scattering length of the bosons vanishes,
as the hard-core repulsion is exactly balanced by the short-range attraction.
Furthermore, the SU$_{\rm C}$(2) symmetry of the Hubbard model implies
that the energy difference per volume of a phase-separated state
(where doubly occupied sites stick together, i.e. where $\vec{S}$ points only up or down)
and a superfluid state vanishes in the thermodynamic limit of the uniform system.

{\em SU$_{\rm C}$(2) symmetry breaking}
will therefore either stabilize the  superfluid state or lead to phase
separation.
Let us list possible corrections to the Hubbard model \eqref{H}
which break the SU$_{\rm C}$(2) charge symmetry (at $V_t=0$),
but preserve the SU(2) spin symmetry.
As the chemical potential is fixed by the particle number, the most important
contributions are two-site terms.
The possible two-site terms are longer-range tunneling, assisted tunneling, pair tunneling,
density-density and spin-spin interactions.
For an optical lattice, where the lattice potential in $x$, $y$, and $z$ directions is independent,
$V(\vec{x})=\sum_i V_i(x_i)$, the leading longer-range tunneling term
is to the second-neighbor site in $x$, $y$ and $z$ direction.
Its strength can be estimated as $J'\sim J^2/\Delta$ where $\Delta$
measures the gap to the next Bloch band of the lattice.
Assisted next-neighbor tunneling of the form
$-\tilde{J}c^\dagger_{i\sigma}c_{j\sigma}(n_{i,-\sigma}+n_{j,-\sigma}-1)$
arises from the interaction correction to the local Wannier wavefunction
of a fermion and is hence given by $\tilde{J}\sim J U/\Delta$.
Finally, both the next-neighbor density-density interaction, $U'(n_i-1)(n_j-1)$, and a next-neighbor pair-hopping
term $J_p c^\dagger_{i \uparrow}c^\dagger_{i \downarrow}c_{j \downarrow}c_{j \uparrow}$ are given
by matrix elements of Wannier states on adjacent sites,
$U'\approx J_p/4 \sim U (J/\Delta)^2$.
A next-neighbor spin-spin interaction is of similar size, but unimportant
for the dynamics of doubly occupied and empty sites.
One may also consider three-site terms, but those are easily seen to be subleading.

After the Schrieffer-Wolff transformation, the leading correction to the dynamics of
double occupancies for $J \ll U \ll \Delta$ arises from the $U'$ repulsion term and the pair hopping $J_p$,
as the tunneling terms only enter in second order via an intermediate singly occupied
state.
In fact, the contributions of $\tilde{J}$ cancel, and the effect of $J'$
is $J'^2/U \sim J^4/(\Delta^2 U)$, such that the effects of $U'$ and $J_p$ are larger
by a factor of $(U/J)^2 \gg 1$.
Hence, in the bosonic language we are left with
\begin{eqnarray}
\Delta \mathcal{H}=  \sum_{\langle ij\rangle}  4 U' (n_{d,i}-\frac1 2) (n_{d,j}-\frac1 2)+J_p d^\dagger_{i} d_j
\end{eqnarray}
and both terms stabilize superfluidity.
Translating into spins and using the same variational Ansatz as above, one obtains the
leading correction to~(\ref{energy}) in $d=3$:
$
\Delta E= (6 U'+3 J_p/2) \int d^3 r \, \hat{n}_z^2
$ where we used $\hat{n}^2=1$.
It is convenient to parameterize the strength of the additional
interactions by the dimensionless parameter $\gamma=(6 U'+3 J_p/2) U
N_d^{2/3}/J^2$. For a typical experimental setup $\gamma \sim
(U/\Delta)^2 N_d^{2/3}$ will be quite large, $10 \lesssim \gamma
\lesssim 1000$.
As shown in Figs. \ref{densityprofile}b and
\ref{mix}b, a large $\gamma$ leads to the expected
expansion of the cloud and therefore to an {\em enhancement} of the
superfluid fraction.

From our variational wavefunction, one can calculate the
momentum distribution of the fermion pairs, $\langle d^\dagger_k d_k
\rangle$. While the non-condensed fraction gives only a smooth
background signal, the condensate gives rise to sharp peaks centered
at $(\pm \pi,\pm \pi,\pm \pi)$, see Fig.~\ref{schematic} and Fig.~\ref{mix}c.

{\em Conclusion.}
We have shown that, within the strongly repulsive Hubbard model, one can
realize a metastable $s$-wave superfluid by expanding a
band-insulating ground state.
One experimental problem may be the preparation of the initial state,
as in present optical-lattice experiments \cite{esslinger08,bloch08} the entropy is not small,
implying a sizeable fraction of singly occupied sites \cite{bloch08} even for
large $V_t$.
Fortunately, the condition for the onset of Bose condensation is not very strict.
For non-interacting bosons, the entropy per particle has to be smaller than $3.6 k_B$.
The corresponding entropy per fermion of $1.8 k_B$
can be reached by cooling non-interacting fermions down to 0.22 $T_F$;
considerably lower temperatures are nowadays obtained routinely \cite{bloch08}.
It is, however, presently not clear how this entropy is distributed
between singly and doubly occupied sites.
Furthermore, it may
not be simple to keep quasi-adiabatic conditions when the cloud is
expanded \cite{esslinger08,bloch08}, as extrinsic heating processes limit the total time in which
experiments are performed.
While a quantitative estimate of these corrections is difficult,
we are optimistic that, with present-day technology and suitably chosen experimental conditions,
an exotic finite-momentum $s$-wave condensate can be realized and detected for strongly
repulsive fermions in optical lattices.


We thank I. Bloch, M. Garst, and U. Schneider for useful discussions,
as well as the DFG (SFB 608 and SFB TR 12)
and the Humboldt foundation for financial support.



\begin{thebibliography}{10}
\bibitem{alternatively}
Alternatively, one can also use a non-interacting band-insulating state as initial configuration.
At $t=0$ the interaction $U \gg D$ is suddenly switched on.
Then the system equilibrates under the condition that $N_d$ is fixed.
For small $N_1/N_d$ the generated entropy remains small.
\bibitem{pot}
It is simple to include the external potential in the algorithm described in \cite{schriefferWolff2}
by adding it to $T_0$, the part of the kinetic energy which does not change $N_d$.
\bibitem{gutzwiller}
Such relations are well-known for Gutzwiller
wavefunctions for hard-core bosons:
Here the Gutzwiller wavefunction is exact for $V_t=0$.
\end{thebibliography}
\end{document}